\documentclass[aps,pre]{revtex4}
\usepackage[fleqn]{amsmath}
\usepackage{graphicx}
\usepackage{textcomp}
\begin{document}
\author{Ilana Bogod}
\email{ibogod@campus.technion.ac.il}
\affiliation{Schulich Faculty of Chemistry, Technion-Israel Institute of Technology, Haifa 32000, Israel}

\author{Saar Rahav}
\email{rahavs@technion.ac.il}
\affiliation{Schulich Faculty of Chemistry, Technion-Israel Institute of Technology, Haifa 32000, Israel}

\title{Kinetic discrimination of a polymerase in the presence of obstacles}

\begin{abstract}
One of the causes of high fidelity of copying in biological systems is kinetic discrimination. In this mechanism larger dissipation and copying velocity result in improved copying accuracy. We consider a model of a polymerase which simultaneously copies a single stranded RNA and opens a single- to double-stranded junction serving as an obstacle. The presence of the obstacle slows down the motor, resulting in a change of its fidelity, which can be used to gain information about the motor and junction dynamics. We find that the motor's fidelity does not depend on details of the motor-junction interaction, such as whether the interaction is passive or active. Analysis of the copying fidelity can still be used as a tool for investigating the junction kinetics.
\end{abstract}

\maketitle
{\setlength{\mathindent}{0.5cm}

\section {Introduction} \label{introduction}
Being alive means being out of thermal equilibrium. Our cells grow and divide via a host of nonequilibrium processes in which complex molecules are synthesized, transported and degraded. Many of these processes are carried out by biomolecules which act like motors or machines, and are driven by chemical potential differences. One of the most delicate and demanding tasks handled by such biological motors is the replication and distribution of genetic information. Polymerase type enzymes, which generate copies of nucleic acid polymers, are a well-known class of information handling molecular machines. Maintaining high fidelity of copying by these enzymes is often crucial, as a large number of copying errors may result in malfunctions and death.
Since “information is physical”, as beautifully stated by Landauer \cite{Landauer1996}, the copying should be studied as a thermodynamic process.

Close to thermal equilibrium, the accuracy of the copying process is determined by the difference in binding free-energy between correct and incorrect monomers.
This mechanism of error reduction has been termed energetic discrimination. But binding free energies of various nucleic acids have a limited range, as they all need to stay bound, yet not be too difficult to remove. This means that energetic discrimination schemes often show only moderate level of accuracy, much lower than what is typically observed in biological systems. The natural conclusion is that accurate copying of information requires out-of-equilibrium processes, as pointed out by Hopfield \cite{Hopfield1974} and Ninio \cite{Ninio1975}. When a system is driven away from equilibrium, copying fidelity can be enhanced at the cost of additional dissipation. A theory of the fidelity of transcription will inevitably investigate such processes from a thermodynamical perspective.

Several recent papers were devoted to this question \cite{Pigolotti2013,Gaspard2008,Gaspard2009,Gaspard2014,Gaspard2016,Gaspard2016a}. Sartori and Pigolotti \cite{Pigolotti2013} discussed the difference between the energetic discrimination scheme that was mentioned above and kinetic discrimination. The latter is controlled by the rates of incorporation of correct and incorrect monomers, or equivalently by activation energies, rather than by differences in binding free-energies. They pointed out that typically one of these discrimination schemes dominates the other, so that the subdominant mechanism has little effect on the copying fidelity. Kinetic discrimination is typically dominant when the process is far from equilibrium.

The thermodynamics of polymerization processes were studied meticulously by Andrieux and Gaspard in a series of papers \cite{Gaspard2008,Gaspard2009,Gaspard2014,Gaspard2016,Gaspard2016a}. For copolymerization on a template with kinetic discrimination, their model showed trade-off between accuracy and dissipation.
Processes driven by larger thermodynamic affinities resulted in faster transcription rates, accompanied by lower error rates.

Such results suggest a connection between the velocity of the polymerizing agent and its accuracy. However, the models studied so far in the context of transcription or copolymerization were allowed to propagate freely on their template. In biological systems, the template -- a single strand of DNA/RNA -- may be blocked by a second strand or a hairpin, which must be removed before copying can proceed. Typically these obstacles are handled by helicases or other specialized molecular machines. Nevertheless, there are known examples of polymerases, such as the T7 RNA polymerase \cite{Richardson1983} and HIV-1 reverse transcriptase \cite{Hottiger1994},  that remove obstacles on their own, simultaneously transcribing and opening the double stranded structure. When reaching an obstacle these polymerases must slow down. How does such an interaction with an obstacle affect the fidelity of transcription? Naively, one would expect that the presence of an obstacle would slow down such a motor and, accordingly, lead to more copying errors.

Betterton and J\"ulicher studied a simple model of a helicase encountering such an obstacle \cite{Julicher2003,Julicher2005}. They qualitatively characterized the interaction between motor and obstacle as being either active or passive. In passive interaction, the motor must wait until the nearest bond in the double-stranded obstacle opens due to a thermal fluctuation, thereby allowing the motor to step forward and prevent the bond from closing. In the active interaction the motor partially enters the obstacle, and the elastic interaction between the motor and the single- to double-strand junction increases the likelihood of bond opening. Active interactions were found to result in higher rates of bond breaking and larger velocities.

In this paper we study how the fidelity of a polymerase is affected when it encounters an obstacle. A simple model which describes both polymerization on a template and interaction with a junction is developed. It combines elements from the models presented by Andrieux and Gaspard \cite{Gaspard2008,Gaspard2009,Gaspard2014}
and by Betterton and J\"ulicher \cite{Julicher2003,Julicher2005}. The model is studied numerically with the help of simulations, and analytically, using a steady-growth ansatz. In particular, differences between copying fidelity of active and passive interactions are investigated.

The structure of the paper is as follows: in section \ref{model} we present a simple model of a polymerase working against an obstacle. We discuss the possible processes and their transition rates. In section \ref{ansatz} we present the master equation that describes the dynamics of the model. We also present a steady-growth ansatz that allows to obtain analytical expressions for observables such as the polymerase mean velocity and copying fidelity. In section \ref{passive} we investigate the case of passive interaction between the polymerase and the obstacle, while section \ref{active} is dedicated to systems with active interactions. In both sections the predictions of the steady-growth ansatz are compared to Monte-Carlo simulations. We find that while the form of the elastic interaction affects the mean copying rate, it has no effect on the copying fidelity. We discuss the implications of this result in section \ref{discussion}.

\section{A Markovian model of a polymerase pushing against an obstacle} \label{model}

\begin{figure}
\includegraphics[width=\linewidth]{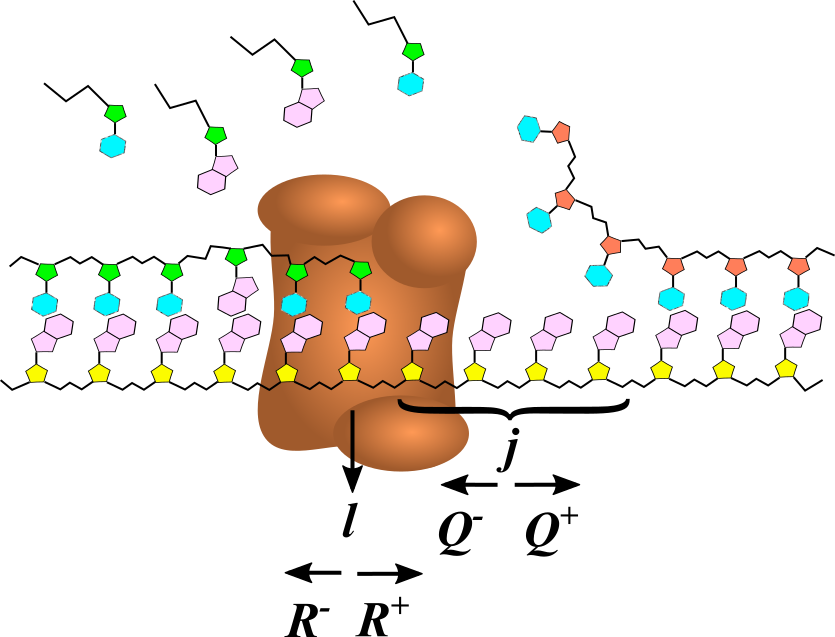}
\caption{A simple model of a polymerase on a single-stranded DNA that serves as a template for copying. The motor is located on the border between a single and double stranded DNA. It can move forward while polymerizing an additional nucleotide onto the complementary strand, or it can move backward while removing a base from this strand. At the same time, the obstacle can move forward, or, if the motor is far enough, the obstacle can move back, as the double strand closes. }
\label{fig:model}
\end{figure}

In this section we present a simple Markovian model of a polymerase. The model is heuristically depicted in Fig. \ref{fig:model}. The system is composed of a substrate nucleic acid polymer (DNA or RNA) with a known  sequence of monomers.  The active site of the polymerase is located at the $l_{th}$ base of the substrate. The polymerase progresses along the chain while adding monomers to the complementary strand, leaving behind a double-stranded structure. The obstacle we study is a junction with another double-stranded structure. These are known to occur as part of a secondary structure of the substrate, such as a hairpin. In the model, the junction is located at site $(l+j)$.

The discreteness of base pairs naturally calls for a model with discrete steps, which are characterized by transition rates.  We consider four different processes:
\begin{enumerate}
 \item Forward motion of the polymerase from site $l$ to site $l+1$, accompanied by addition of a monomer $m_{l+1}$ to the complementary strand. The rate of this process is denoted by $R^+_{m_{l+1}}$.
 \item Backward motion of the polymerase from site $l$ to site $l-1$, while removing monomer $m_l$ from the complementary strand. The rate is denoted by $R^-_{m_l}$.
 \item Opening of a bond in the junction, with rate $Q^+$. In this process the junction moves forward along the template.
 \item Closing of a bond in the junction, with rate $Q^-$, moving the junction backwards.
\end{enumerate}

An important distinction between the addition and removal of monomers is that the polymerase can always add any of the monomers in the solution, but can remove only the specific monomer which is at the terminal position on the complementary strand. The kinetic equations that we will use in the next section will take this into account.

The polymerase and junction will interact with each other if they are close enough. Physically, this is an elastic interaction that is caused by deformation of the junction when the polymerase pushes into it. Such an interaction will partially destabilize the outer bond in the junction while pushing the polymerase backwards. We assume that the interaction changes the transition rates according to  $R^+_m(j)=r^+_m \Theta ^-(j)$,  $R^-_m(j)=r^-_m \Theta ^+(j)$,  $Q^+(j)=q^+ \Theta ^+(j)$ and  $Q^-(j)=q^- \Theta ^-(j)$. $r ^\pm _m$ and $q^\pm$ are the rates for the non-interacting (or distant) polymerase and junction. $\Theta^\pm (j)$ describe the polymerase-junction interaction. A more detailed description of the dependence of these rates on system properties is given in the rest of this section.

We wish to study the fidelity of copying and how it is affected by the interaction with the junction. Two qualitatively different mechanisms can be employed to improve fidelity of copying. Energetic discrimination favors the correct monomer based on lower free energy of binding. In contrast, kinetic discrimination originates from different kinetic rates for binding of monomers. As pointed out by Sartori and Pigolotti \cite{Pigolotti2013}, the two mechanisms compete with each other, since energetic discrimination works near equlibrium, while its kinetic counterpart reaches best fidelity far from equilibrium.

Here we consider a model with pure kinetic discrimination, following a similar choice by Bennett \cite{Bennett1982} and by Andrieux and Gaspard \cite{Gaspard2008}.  We make several simplifying assumptions about the structure of the model. These assumptions allow to reduce the bookkeeping involved in defining the system's state, while keeping the main qualitative features of the dynamics. Specifically, we assume that the substrate strand is built out of two types of monomers and the solution has the two complementary monomers in equal  concentrations and in a spatially homogeneous mixture. With these assumptions, the transition rates only discriminate between addition of a correct or an incorrect monomer to the complementary strand, where the correctness is determined by comparing the monomer to its partner on the substrate. Crucially, there is no need to specify the composition of the substrate.

The absence of energetic discrimination means that
\begin{equation}
\label{eq:deltaG}
\frac{r^+_m}  {r^-_m} = \frac{[mNTP] \hat r^+_m}{[PPi] \hat r^-_m}  =  \frac{[mNTP]}{[PPi]} \epsilon ,
\end{equation}
for any value of $m$, where $\epsilon = \exp \left[ -  \Delta G^0 \right]$, $\Delta G^0$ is the standard free energy of the polymerization reaction in units of $k_B T$, 
$[mNTP]$ is the concentrations of nucleotide $m$ and $[PPi]$ is the concentration of pyrophosphate, a byproduct of the polymerization reaction. 
$\hat r ^\pm _m $ denotes the transition rates to addition and removal of monomers at standard concentrations. We assume that concentrations do not vary in time. This is a good approximation for many \textit{in vitro} experiments. 

Kinetic discrimination can be expressed through higher rates of addition and removal of the correct monomer $(c)$, compared to the wrong monomer $(w)$, namely
\begin{equation}
\label{eq:define_d}
\frac{\hat r^+_c}  {\hat r^+_w} =\frac{\hat r^-_c}  {\hat r^-_w} = d,
\end{equation}
where $d$ parametrizes the preference for inserting correct monomers, and therefore the resulting fidelity of copying. In living cells $d \approx 10^4-10^6$ \cite{LJ2006}, but we will consider smaller values to avoid problems associated with insufficient sampling of errors in our simulations. Kinetic discrimination reduces errors because when the polymer grows rapidly, a larger proportion of the incorporated monomers is of the correct type. These monomers are left behind in the copied strand when the polymerase propagates. They are bound from both sides, and are unlikely to detach unless the motor performs multiple backward steps to return to their location in the chain.

Based on these considerations, the transition rates for pure kinetic discrimination will take the following form
\begin{equation}
\label{eq:def_r}
\begin{split}
&r^-_w=\hat{r}^-_w[PPi], \\
&r^-_c=\hat{r}^-_c[PPi]=\hat{r}^-_w d [PPi], \\
&r^+_w=\hat{r}^+_w[wNTP]=\hat{r}^-_w \epsilon [wNTP], \\
&r^+_c=\hat{r}^+_c[cNTP]=\hat{r}^-_w d \epsilon [cNTP], \\
\end{split}
\end{equation}

The opening and closing of the bonds in the junction are ruled by thermal fluctuations. The ratio of rates has been determined empirically to be $ \frac{q^-}  {q^+} \approx 7$  \cite{Lohman1996,Julicher1998}. This expresses the fact that in equilibrium the two strands of the junction tend to bind to create a double-stranded structure. The motor must be driven against the junction to reverse this trend.

Finally, the elastic interaction effect on the rates enters through the factors $ \Theta ^ \pm $. Thermodynamic consistency mandates that
\begin{equation}
\label{eq:theta_ratio}
\frac{\Theta ^+ (j)}  {\Theta ^- (j+1)} = e ^{\left[ U(j) - U(j+1) \right]} ,
\end{equation}
where $U(j)$ is the potential of the interaction at distance $j$ in units of $k_BT$. We will assume that the influence of this elastic interaction affects both rates according to
\begin{equation}
\label{eq:Theta}
\begin{split}
\Theta ^+(j) = e^{(g-1)[U(j+1)-U(j)]} ,\\
\Theta ^-(j+1) = e^{g[U(j+1)-U(j)]} . \\
\end{split}
\end{equation}
Here, $g$ is a load-distribution-factor like parameter. Note that the interaction affects both processes that close the distance between the motor and the junction in the same way, and the same is true for opening. It does not matter if the process involves motion of the motor accompanied by a polymerization reaction, or motion of the junction due to the formation of a new inter-strand bond. The same form of interaction was also used by Betterton and J\"ulicher in their model of helicase \cite{Julicher2003,Julicher2005}.

\section{The master equation and the steady-growth ansatz} \label{ansatz}

Consider a system in which the polymerase was placed on the substrate strand and then left to evolve under conditions of mean growth. The state of the system is characterized by: i) the position of the motor $l$ on the substrate, which is
assumed to be initially at $l=0$. $l$ is therefore also the length of the complementary strand being polymerized; ii) the composition of the complementary strand compared to the substrate strand: $m_1, m_2,...m_l$, where $m_k=c$ or $w$; and iii) the distance of the motor from the junction $j$.

Given the processes and rates described in Sec. \ref{model}, the probability distribution of the system evolves according to the master equation
\begin{equation}
\label{eq:master_1}
\begin{split}
& \frac{dP}{dt}(m_1...m_l,j,l,t)   = \\
& \indent R^+_{m_{l}} (j+1) P(m_1...m_{l-1},l-1,j+1,t) + \sum _{m_{l+1}} R^-_{m_{l+1}} (j-1) P(m_1...m_{l+1},l+1,j-1,t) \\
& \indent + Q^+(j-1) P(m_1...m_{l},l,j-1,t) + Q^-(j+1) P(m_1...m_{l},l,j+1,t) \\
& \indent-\left[  Q^+(j)+Q^-(j)+R^-_{m_{l}}(j) + \sum _{m_{l+1}} R^+ _{m_{l+1}}(j) \right]  P(m_1...m_{l},l,j,t).
\end{split}
\end{equation}

The system evolution can be studied in full detail by solving the master equation with an appropriately chosen initial condition, or alternatively by simulating the underlying jump process. However, both approaches are needlessly complicated if one is interested in simple quantities such as the mean error rate and velocity.

A simpler approach for the description of these observables, which may even allow for an analytical solution, was developed by Andrieux and Gaspard \cite{Gaspard2008}. The approach is based on the assumption that after a transient, the system reaches a steady-growth regime in which correlations between the length of the chain $l$, the composition of the chain and the distance $j$ are lost. In this steady-growth regime the probability distribution can be approximated by
\begin{equation}
\label{eq:ansatz}
P(m_1...m_l, l, j, t) \simeq P_t(l)  \mu (m_1...m_l) \Phi(j),
\end{equation}
where $P_t(l)$ is the probability of length $l$, $\mu (m_1...m_l)$ is the probability of a given sequence when the length is set, and $\Phi(j)$ is the probability of distance $j$. Only the length distribution is explicitly time-dependent.

This ansatz can not be exact, since the distribution $\mu (m_1...m_l)$ must include the memory of transient behavior in the distant past. Nevertheless, it is a useful approximation in the steady-growth regime, as long as one focuses on marginal distributions such as $ \mu(m_l), \mu(m_{l-1},m_l) $ etc., describing the probability distribution of one or a few monomers near the tip of the growing strand. In the steady-growth regime one expects these distributions to be independent of time and the precise value of  $l$. Using similar physical intuition, one expects the probability $ \Phi (j) $ to reach a time-independent steady state in the steady-growth regime.  The equations describing these simpler marginal probabilities are derived by substituting the ansatz Eq. (\ref{eq:ansatz}) into the master equation and summing over all unwanted variables.

Summation over the chain composition and over values of $j$ gives an equation for the length distribution
\begin{multline}
\label{eq:speed1}
 \frac{dP_t(l)}{dt} =
\sum _{m_{l}}r^+_{m_l} \left\langle \Theta ^- \right\rangle P_t(l-1)
+ \sum _{m_{l+1}}r^-_{m_{l+1}} \left\langle \Theta ^+ \right\rangle \mu (m_{l+1}) P_t(l+1)
\\
- \left[\sum _{m_{l+1}}r^+_{m_{l+1}} \left\langle \Theta ^- \right\rangle
+ \sum _{m_{l}}r^-_{m_{l}} \left\langle \Theta ^+ \right\rangle \mu (m_{l}) \right] P_t(l).
\end{multline}
\underline{Here} $ \left\langle \Theta ^\pm \right\rangle = \sum _j \Theta ^\pm (j) \Phi (j) $, and  $ \mu (m_l) $ is the likelihood that the last monomer in the chain is $m_l$.  We will see shortly that for our model there are no correlations in the composition of the chain and so $ \mu (m_l) $ can have the two values $ \mu (w) $ and $ \mu(c)=1-\mu(w)$.
$ \mu(w) $ is also the probability of copying error in the bulk of the copied strand.

Inspection of Eq. (\ref{eq:speed1}) reveals that it includes processes in which the chain grows and shrinks. The mean growth velocity is the difference between the mean rate of polymerization and depolymerization, namely
\begin{equation}
\label{eq:speed2}
v = \sum _m r^+_m \left\langle \Theta ^- \right\rangle  - \sum _m r^-_m \mu(m) \left\langle  \Theta ^+ \right\rangle .
\end{equation}

Summation over the chain composition and the lengths leads to an equation for the distribution of distances between the motor and junction.  A short calculation gives
\begin{multline}
\label{eq:phi}
\big(  \sum _m r^+ _m + q^- \big)   \Theta ^- (j+1) \Phi (j+1) +
\big( { \left\langle r^- \right\rangle + q^+} \big) \Theta ^+ (j-1) \Phi (j-1) \\
\indent -\left[ \big(\sum _m r^+ _m +q^- \big) \Theta ^- (j)
+ \big( \left\langle r^- \right\rangle + q^+ \big) \Theta ^+ (j)\right]  \Phi (j)=0
\end{multline}
as the equation determining the distribution of $ \Phi (j) $. Here $ \left\langle r^- \right\rangle = \sum _m r^- _m \mu (m) $ is the mean rate of removal of the last monomer.

To obtain equations for the composition of the monomers in the complementary strand, one sums over all values of $l$, $j$ and
the possible composition of the first $k$ monomers
$m_1 ... m_k$. This results in a hierarchy of equations for
the probability distribution of the last few monomers. The first two equations in this hierarchy are given by
{\setlength{\mathindent}{0cm}
\begin{equation}
\label{eq:h1}
 r^+ _{m_l} \left\langle \Theta ^- \right\rangle
+ \sum _{m_{l+1}} r^-_{m_{l+1}}\left\langle \Theta^+ \right\rangle \mu (m_l m_{l+1}) - r^- _{m_l} \left\langle \Theta ^+ \right\rangle \mu (m_l)
- \sum _{m_{l+1}} r^+ _{m_{l+1}} \left\langle  \Theta ^- \right\rangle \mu (m_l) =0 ,
\end{equation}
\begin{equation}
\label{eq:h2}
r^+ _{m_l} \left\langle \Theta ^- \right\rangle \mu(m_{l-1})
+ \sum _{m_{l+1}} r^- _{m_{l+1}} \left\langle \Theta^+ \right\rangle \mu (m_{l-1} m_l m_{l+1}) - r^- _{m_l} \left\langle \Theta ^+ \right\rangle \mu (m_{l-1}m_l)
- \sum _{m_{l+1}} r^+ _{m_{l+1}} \left\langle  \Theta ^- \right\rangle \mu (m_{l-1} m_l )=0 .
\end{equation}
{\setlength{\mathindent}{1cm}
This hierarchy has a solution in which there are no correlations between consecutive monomers. 
One can show that under the assumption of correlations only between nearest neighbors, which allows closing the hierarchy using Eqs. (\ref{eq:h1}) and (\ref{eq:h2}), the conditional
probability $\mu (m_{l-1}|m_l)$ tends to $\mu(m_l)$. More importantly, in Secs. \ref{passive} and \ref{active} we compare the results of the ansatz
to simulations which do not assume lack of correlations. Excellent agreement is found. Both facts strongly suggest that the uncorrelated solution
is stable to small perturbations.

Since there are no correlations in the steady-growth regime, $ \mu (m_l)$ describes also the probability to find a monomer anywhere on the chain , and $ \mu(m_1m_2...m_l)= \mu (m_1) \mu(m_2)...\mu (m_l)$. We note in passing that models with rates leading to nearest-neighbor correlations were studied by Andrieux and Gaspard \cite{Gaspard2014,Gaspard2016,Gaspard2016a}.
In absence of correlations one obtains the following set of equations for monomer probabilities
\begin{equation}
\label{eq:probabilities}
r^+ _{m_l} \left\langle  \Theta ^- \right\rangle
+ \sum _{m_{l+1}} r^-_{m_{l+1}}\left\langle \Theta ^+ \right\rangle \mu (m_l) \mu (m_{l+1})
- \left[ r^-_{m_l} \left\langle  \Theta ^+ \right\rangle
+ \sum _{m_{l+1}} r^+ _{m_{l+1}}\left\langle \Theta ^- \right\rangle \right]  \mu (m_l) = 0 .
\end{equation}
For the model studied here, the monomers in the complementary strand can either match the substrate, $m=c$, or be a copying error, $m=w$. By substituting $ \mu (c) = 1 - \mu (w) $
one can reduce Eq. (\ref{eq:probabilities}) to a single equation for the copying fidelity
\begin{equation}
\label{eq:err_rate}
\left\langle \Theta ^+ \right\rangle (r^- _w - r ^- _c) \mu ^2 (w)
- \left[(r^- _w - r ^- _c)\left\langle \Theta ^+ \right\rangle
+ (r^+ _w + r ^+ _c)\left\langle \Theta ^- \right\rangle \right] \mu (w)
+ r^+ _w \left\langle \Theta ^- \right\rangle = 0.
\end{equation}

Equations (\ref{eq:speed2}), (\ref{eq:phi}) and (\ref{eq:err_rate})
are a set of coupled equations that characterize the properties of the polymerase in the steady-growth regime. Their simple form allows one to calculate quantities such as the motor's mean velocity and its fidelity analytically. The simple form of these equations ultimately emerges from the simple dependence
of the transition rates on the chain composition and the motor-junction distance. In the next two sections  we will solve these equations explicitly for two cases. We will also compare their predictions to those of a stochastic simulation which does not assume a steady-growth regime.

\section {Passive unwinding} \label{passive}

Following Betterton and J\"ulicher \cite{Julicher2003,Julicher2005} we qualitatively characterize the interaction between the polymerase and the junction as either passive or active. In
the passive case, the interaction between motor and junction is that of a hard wall. As a result, the motor does not enter the junction, and equivalently, a transition that closes the junction on the motor is impossible.  Active unwinding, which will be studied in the next section, allows the motor to enter the junction. As will be seen later, this modifies the transition rates in a way which can result in faster unwinding.

The hard wall interaction means that $ \Phi (j) = 0 $ for  $ j \le 0 $. In addition, the transition from $j=1$ to $j=0$ is forbidden, so $ \Theta ^- (1) = 0$. When the motor is away from the junction, its interaction with the wall can be neglected, and $ \Theta ^- (j) = 1$ for $ j > 1$, while $ \Theta ^+ (j) =1$ for all values of $j$. This interaction is termed passive since unwinding happens when a bond in the junction opens due to a purely thermal fluctuation and the motor steps into the newly available space, thereby preventing the bond from closing.
When this rectification process is more likely than its reversed process, the double-stranded DNA/RNA junction will be unwound on average. An inspection of Eq. (\ref{eq:phi}), shows that $ \Phi (j) $ must in fact satisfy a detailed balance condition
\begin{equation}
\label{eq:balance}
\big[r^+ _w + r^+ _c + q^- \big] \Theta ^- (j+1) \Phi (j+1) =
\big[ q^+ + \left\langle  r^- \right\rangle \big]
\Theta ^+ (j) \Phi (j) .
\end{equation}
The underlying reason for the appearance of this detailed balance condition is the one-dimensional structure of the states and transition topology in the variable $j$, which precludes non-trivial closed cycles of transitions.

Substitution of $ \Theta ^\pm (j) = 1 $ for $j>1$, and of $ \Theta ^+ (1) = 1, \Theta ^- (1) = 0$, leads to
\begin{equation}
\label{eq:flux_rho}
\Phi (j+1) = \rho \Phi (j),
\end{equation}
with $ \rho = \frac {\left\langle r^- \right\rangle + q^+ }{r^+ _w + r^+ _c + q^-}$.
We are interested in systems in which the driving force is sufficient for polymerization in absence of a junction, while the junction tends to close in absence of a polymerase. Under such conditions $\rho < 1$. This allows us to explicitly solve for $ \Phi (j) $ as a function of the mean fidelity of copying. We find that $\Phi (j) = (1-\rho)\rho^{j-1}$ for $j \ge 1 $, and $ \Phi (j)=0$ otherwise.
\\
This local-equilibrium distribution $ \Phi(j)$ allows us to calculate the averages
\begin{equation}
\label{eq:avg_Theta_p}
\left\langle \Theta ^+ \right\rangle = \sum _j \Theta ^+ (j) \Phi (j) = 1,
\end{equation}
\begin{equation}
\label{eq:avg_Theta_m}
\left\langle \Theta ^- \right\rangle = \sum _j \Theta ^- (j) \Phi (j)
= \sum ^\infty _{j=2} \Phi (j) = 1 - \Phi (1) = \rho.
\end{equation}

Substituting Eqs.(\ref{eq:avg_Theta_p}) and (\ref{eq:avg_Theta_m}) in Eq. (\ref{eq:err_rate}) gives the following quadratic equation
for the probability of making a copying error
\begin{equation}
\label{eq:err_rate2}
\mu ^2 (w) q^- (r ^- _c - r^- _w)
+ \mu (w) [q^+(r^+ _w + r ^+ _c)- q^- (r ^- _c - r^- _w)+r^-_w r^+_c + r^-_c r^+ _w]
- r^+ _w(r^-_c + q^+)= 0.
\end{equation}

Let us first examine the case of a fixed and immobile junction. In this case the polymerase will not be able to propagate at all, and kinetic discrimination is not possible. By substituting $q^\pm =0$ into Eq. (\ref{eq:err_rate2}) we find that the copying fidelity is
\begin{equation}
\label{eq:equilibrium}
\mu(w) = \frac {r^+_w r^-_c}{r^+_cr^-_w+r^-_cr^+_w}=\frac{1}{2},
\end{equation}
which is the equilibrium error rate for our model due to the equal binding free-energies of $m=w,c$ \underline{(cf. Eq. \ref{eq:def_r})}.

For the general case of $q^-,q^+ \ne 0$ we obtain
\begin{equation}
\label{eq:err_rate3}
\mu(w) =
\frac
{-z + \sqrt{z^2
+4q^- (r ^- _c - r^- _w)r^+ _w(r^-_c+q^+)}}
{2q^- (r ^- _c - r^- _w)},
\end{equation}
with $z=q^+(r^+ _w + r ^+ _c)- q^- (r ^- _c - r^- _w)+r^-_w r^+_c + r^-_c r^+ _w$.
The motor's mean velocity can be calculated by substituting the solutions for $\mu(m)$ and $ \Theta (j) $ in Eq. (\ref{eq:speed2}). A short calculation gives
\begin{equation}
\label{eq:speed3}
v = q^+ -q^- \rho = q^+ - q^- \frac
{\left\langle r^- \right\rangle +q^+}
{r^+_w+r^+_c+q^-} .
\end{equation}

We see that the bond opening rate $q^+$ bounds the possible velocity of the polymerase. This bound is achieved in the limit of high concentrations, where $r^+_w,r^+_c \gg \left\langle r^- \right\rangle , q^\pm$. In this limit, the motor is very likely to reside near the junction and almost immediately step forward once a bond in the junction has opened, making the bond opening in the junction the rate-limiting process.

The analytical calculation leading to Eqs. (\ref{eq:err_rate3}) and (\ref{eq:speed3}) for the copying fidelity and velocity is based on the steady-growth assumption. To test the validity of this assumption, we compared the resulting prediction of the theory to a stochastic simulation of the system. The simulation employed the Gillespie algorithm to determine the next step \cite{Gillespie1976}. The copying fidelity was calculated by counting the fraction of wrong monomers along the chain, while ignoring an initial transient of length 100 base pairs. To gather enough statistics we repeated each simulation 3000 times.
\begin{flushleft}

\begin{figure}[h]
\begin{minipage}{0.49\linewidth}
  \includegraphics[width=\linewidth]{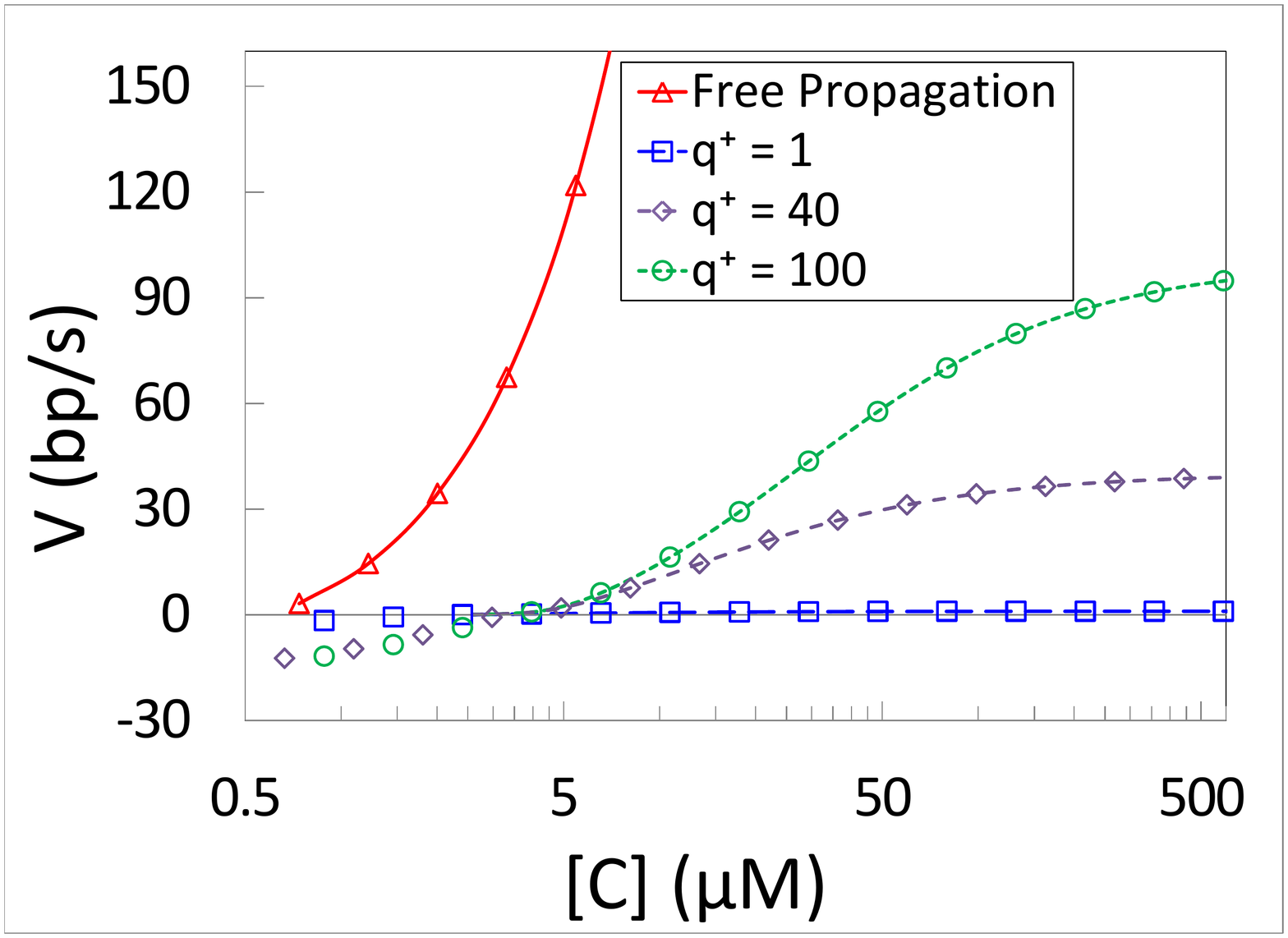}
(a)
\end{minipage}%
\begin{minipage}{.49\textwidth}
  \includegraphics[width=\linewidth]{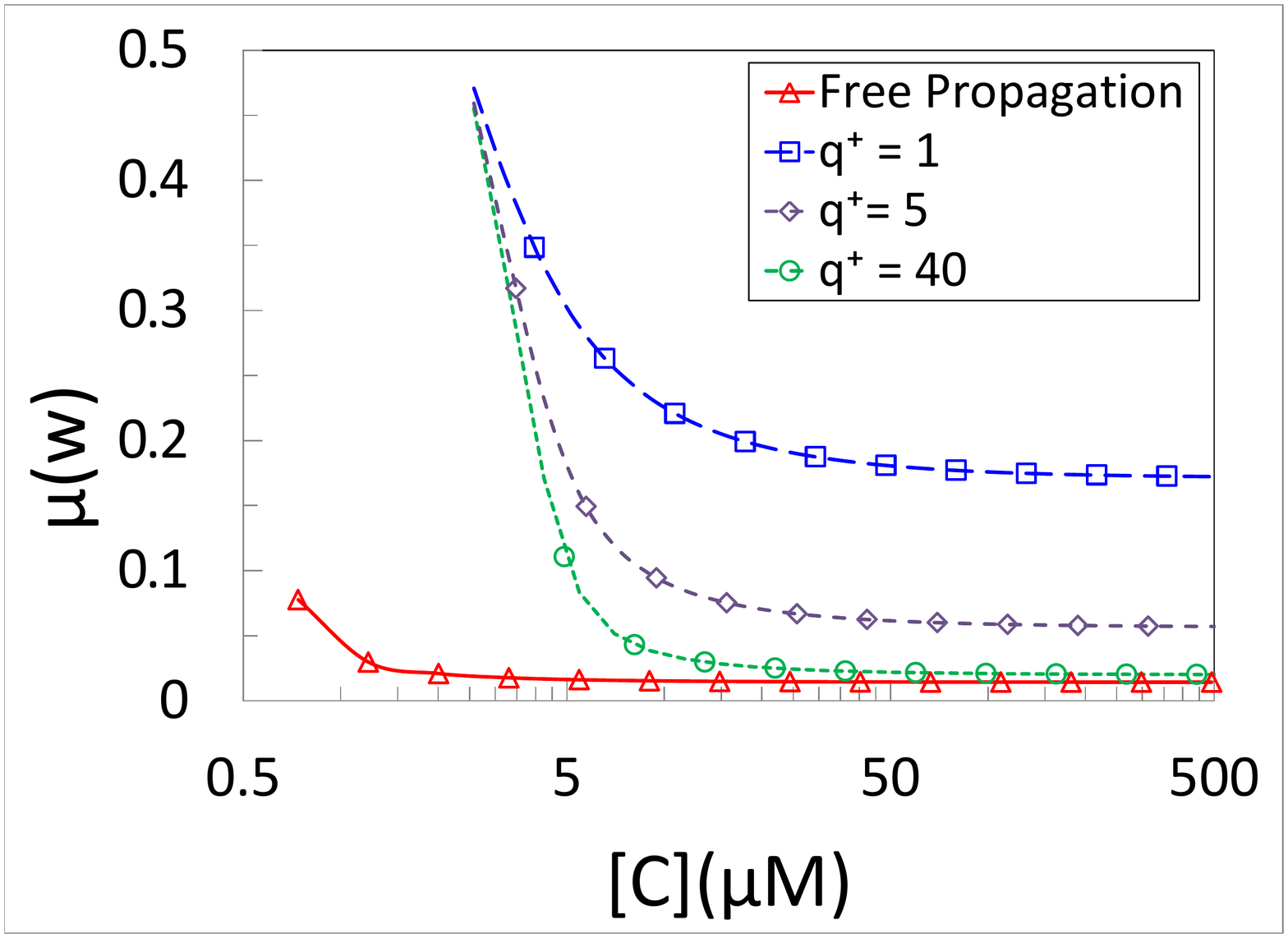}
(b)
\end{minipage}
\caption{(a)
Mean polymerization velocity as a function of monomer concentration [C].
(b) The copying fidelity of the polymerase as a function of monomer concentration.
In both panels, the lines represent the theoretical prediction, obtained from the steady-growth ansatz. Symbols represent the results from a Monte Carlo simulation of the model. }
\label{fig:HW_results}
\end{figure}
\end{flushleft}
Fig. \ref{fig:HW_results}(a) depicts results for the mean velocity of polymerization as a function of monomer concentration. In all our simulations [cNTP]=[wNTP]=[C]. The concentration of PPi was taken to be $100 \mu M$. The kinetic discrimination parameter was chosen to be $d=70$. This value is much smaller than what one expects to find in biological systems. On the other hand, it results in enough copying errors to allow comparison of simulation and theory on a reasonable time scale. Lines correspond to the prediction of Eqs. (\ref{eq:err_rate3}) and (\ref{eq:speed3}), while symbols are the simulation results. It is clear that the agreement is excellent for all positive velocities. Different curves correspond to different values of the kinetic coefficient of the junction $q^+$, where we always keep $q^- = 7q^+$ \cite{Lohman1996,Julicher1998}.

The results clearly show that the polymerase's mean velocity increases with monomer concentration. For a freely propagating motor, the velocity is asymptotically linear in the concentrations, due to the linear dependence of $r^+_{w,c}$. This is no longer true when the junction is present, since the velocity approaches $q^+$ in this limit.
For low concentrations of monomers, the mean velocity is negative and the complementary strand is degraded by the motor. This violates the assumptions made for the steady-growth regime, as the chain composition
was mostly determined by the process used to prepare it, rather than by the polymerase dynamics. Andrieux and Gaspard \cite{Gaspard2014,Gaspard2016} discussed the velocity in a depolymerization regime in detail, but such a discussion goes beyond the scope of the current work.

The copying fidelity is depicted in Fig. \ref{fig:HW_results}(b). It increases with the concentration, as expected in a kinetic discrimination mechanism. The presence of the obstacle reduces the copying fidelity. At slow velocities, the copying error probabilities approach $ \frac{1}{2} $, as predicted. At large concentrations the error probability approaches the value
\begin{equation}
\label{largeLim}
\mu (w) \simeq \frac{r^-_c + q^+}{q^+(1 + d) +2r^-_c}.
\end{equation}
It is easy to see that when $ q^+ \gg r^- $, the obstacle opens fast enough to allow almost free propagation of the motor, and the error rate goes to $\frac{1}{1+d}$
which is the error rate of a far-from-equilibrium freely propagating polymerase \cite{Gaspard2008}. When $ q^+ \ll r^- $, the obstacle is essentially immobile, and the error rate approaches $ \frac{1}{2} $, as expected.

\section {Active unwinding} \label{active}
Active unwinding occurs when the polymerase can push into the junction and drive the two strands apart. The elastic interaction between motor and junction weakens the bond between the strands while also applying a force that pushes the polymerase away from the junction. The precise form of the interaction is not known. We will therefore choose an interaction which exhibits all the expected qualitative features but is easy to use in calculations.

Following Betterton and J\"ulicher \cite{Julicher2003,Julicher2005}, we consider a step-like interaction potential $ U(j) $. The potential is schematically depicted in Fig. \ref{fig:potential}. This potential vanishes when the polymerase is away from the junction ($ j > 1$). When the polymerase enters the junction ($j=0$), the potential obtains the value $U_0 > 0$, expressing a repulsive interaction between motor and junction. The polymerase can push itself further into the junction, where in every step the potential increases by an additional $U_0$. We assume that the polymerase can at most penetrate a finite number of steps into the junction. This is expressed by placing a hard wall interaction at $j=-n$.
\begin{figure}
\includegraphics[width=0.7\linewidth]{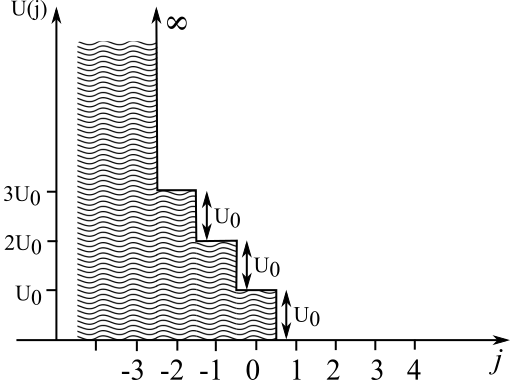}
\caption{
Schematic drawing of the potential of the elastic interaction between the polymerase and the junction as a function of the distance between them. }
\label{fig:potential}
\end{figure}
This potential enters the kinetic equation through the factors $ \Theta ^\pm (j)$ given in Eq. (\ref{eq:Theta}). For the step potential, these factors obtain a simple form
\begin{equation}
\Theta ^+(j)= \begin{cases}
Y^g & ,  j < 1\\
1 & , j \ge 1,
\end{cases}
\end{equation}
and
\begin{equation}
\Theta ^-(j)= \begin{cases}
0&,  j \le -n\\
Y^{g-1}&,  -n < j \le 1\\
1&, j > 1.
\end{cases}
\end{equation}
where $Y \equiv \exp(U_0)$.

The calculation that allowed for an explicit solution of Eq. (\ref{eq:phi}) for the passive case can also be applied for active interactions with this staircase potential. The solution of Eq. (\ref{eq:phi}) follows from the detailed balance condition (\ref{eq:balance}), but the different form of $ \Theta ^\pm $ results in a somewhat different recursion relation for the probability distribution $ \Phi(j) $. For $ j\ge 1$, one still has $ \Phi(j+1)= \rho \Phi(j)$ as in Sec. \ref{passive}, but now for $-n<j<1$ we have $ \Phi(j+1) = \Phi(j) \rho Y $, and $\Phi(j)=0$ for $j \le -n $. A straight forward, but somewhat tedious, calculation gives
\begin{equation}
\label{eq:phi_1}
\Phi(j)= \begin{cases}
\frac{(1-\rho)(1-Y\rho)}{(Y \rho)^{-n}(1-\rho)+\rho(1-Y)}&, j=1\\
\Phi(1) \rho ^ {j-1}&, j>1\\
\Phi(1) (\rho Y)^{j-1}&, -n<j<1 \\
0 &, j \le -n .
\end{cases}
\end{equation}
This can be used to calculate
\begin{equation*}
\left\langle \Theta ^+ \right\rangle=
\sum \limits _{j=1} ^{\infty} \Phi (j)+\sum \limits _{j=1-n} ^{0} \Phi (j)Y^g=
\frac{1-Y \rho +(1-\rho)[(Y\rho)^{-n}-1]Y^g}
{\rho(1-Y)+(1-\rho)(Y\rho)^{-n}} \quad,
\end{equation*}
and
\begin{equation*}
\left\langle \Theta ^- \right\rangle=
\sum \limits _{j=2} ^{\infty} \Phi (j)+\sum \limits _{j=2-n} ^{1} \Phi (j)Y^{g-1}=
\frac{\rho(1-Y  \rho) +(1-\rho)[(Y\rho)^{-n+1}-Y\rho]Y^{g-1}}
{\rho(1-Y)+(1-\rho)(Y\rho)^{-n}}
\quad .
\end{equation*}
\\
The factors $\left\langle \Theta ^+ \right\rangle $ and $\left\langle \Theta ^- \right\rangle $ are subsequently used in the calculation of the polymerase velocity and fidelity.

Interestingly, we note that $\frac{\left\langle \Theta ^- \right\rangle}{\left\langle \Theta ^+ \right\rangle}=\rho$, exactly as in the passive case. Since the mean error rate depends only on this ratio, we find that the motor's fidelity is independent of the elastic interaction and is given be Eq. (\ref{eq:err_rate3}).
The motor's velocity does depend on the elastic interaction and is given by
\begin{equation}
\label{eq:speed_ac}
v=\frac
{\left[ 1-Y \rho +(1-\rho)\left( (Y\rho)^{-n}-1\right) Y^g\right]
\left[ \left\langle r ^- \right\rangle + \rho (r^+_w+r^+_c)\right] }
{\rho(1-Y)+(1-\rho)(Y\rho)^{-n}}
\end{equation}

\begin{figure}[h]
\begin{minipage}{.49\textwidth}
  \includegraphics[width=\linewidth]{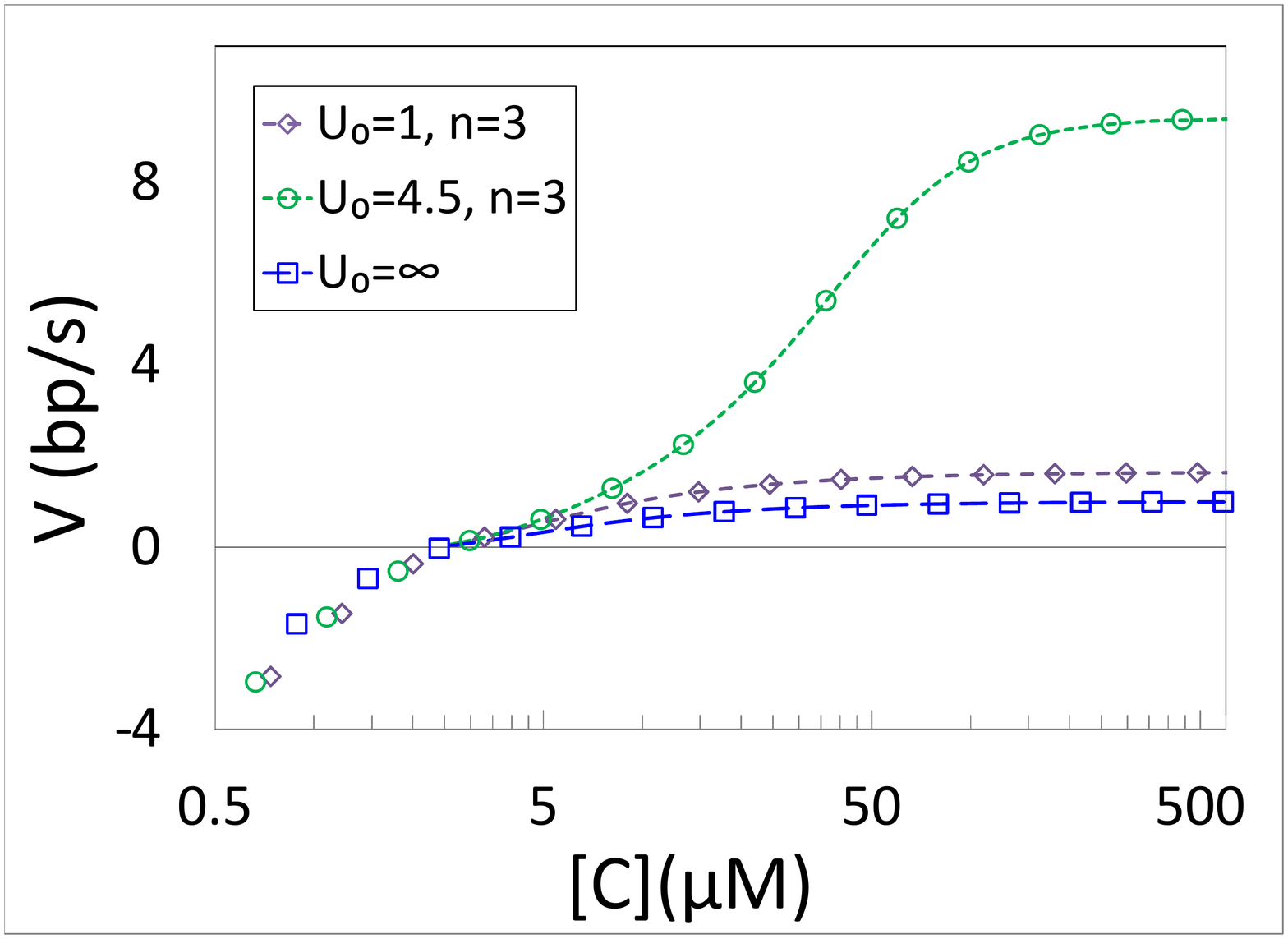}
  (a)
\end{minipage}%
\begin{minipage}{.49\textwidth}
\includegraphics[width=\linewidth]{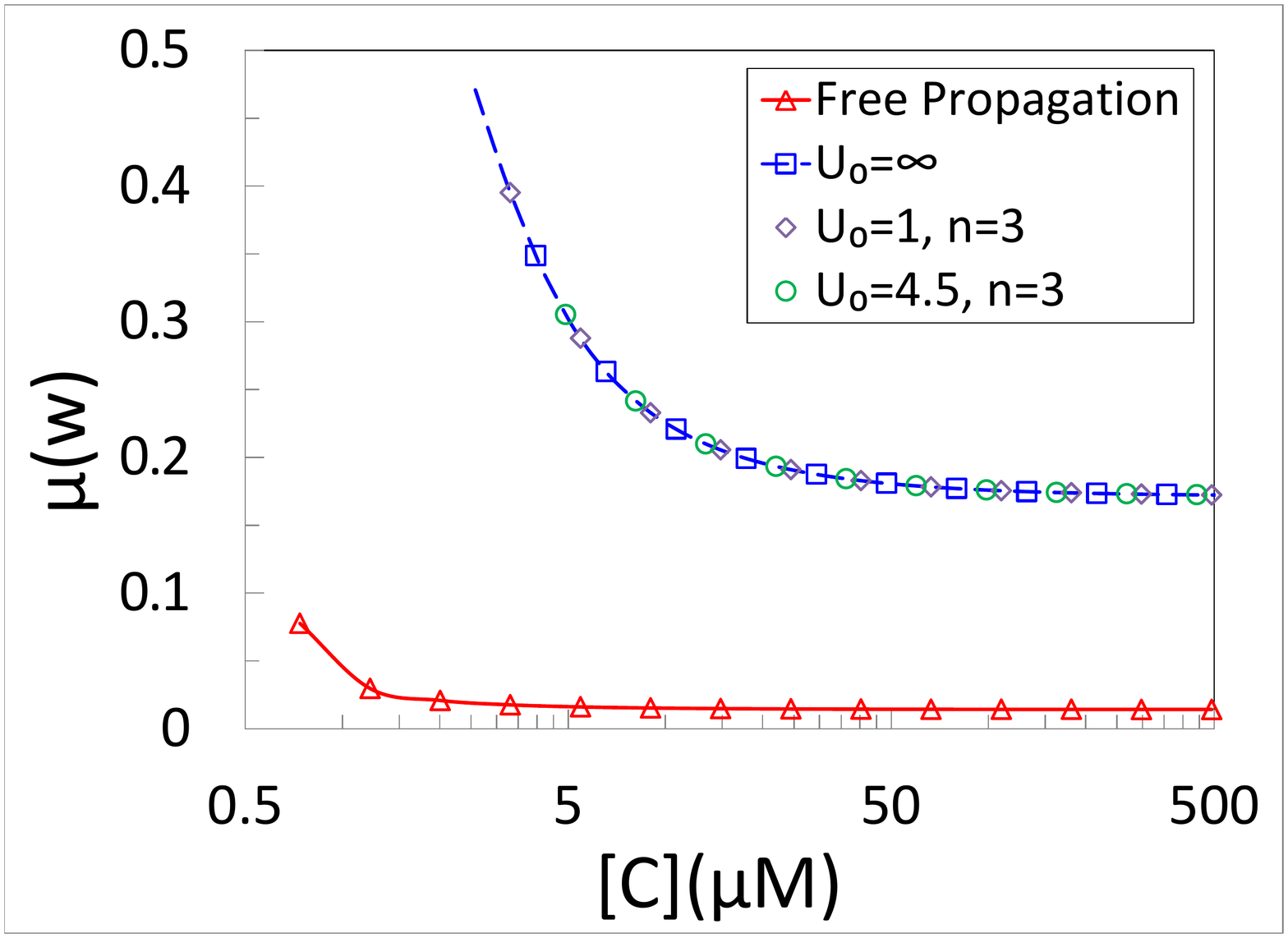}
(b)
\end{minipage}
\caption{(a)  Mean polymerization velocity as a function of monomer concentration $[C]$. Different curves represent different motor-junction interaction parameters, while kinetic parameters of the junction are kept constant ($q^+=1$). (b) The copying fidelity of the polymerase as a function of monomer concentration. Symbols correspond to results from stochastic simulations, whereas lines  depict analytical results. The collapse of three lines onto a single curve shows that details of the interaction have no effect on the fidelity, namely that the error rate for any value of $U_0$ and $n$ are the same as for the hard wall interaction. The solid line and the triangle symbols correspond to a polymerase without an obstacle. Such a freely propagating polymerase exhibits considerably higher fidelity.}
\label{fig:steps_results}
\end{figure}
The analytical predictions of Eqs. (\ref{eq:err_rate3}) and (\ref{eq:speed_ac}) were compared to simulations in Fig. \ref{fig:steps_results}.
Figure \ref{fig:steps_results}(a) shows comparison of the mean velocity of polymerization as a function of monomer concentration for various values of $U_0$. 
The $U_0 \rightarrow \infty$ results correspond to a passive motor with a
hard wall interaction. 
The concentration of PPi and the kinetic discrimination are as in the passive unwinding. Lines, again, correspond to the prediction of Eq. (\ref{eq:speed_ac}),  while symbols are the simulation results. The kinetic coefficients of the junction were set to $q^+=1,q^-=7$ for all the results depicted in the figure.
Excellent agreement is found between Eq. (\ref{eq:speed_ac}) and the simulation results. The polymerization velocity clearly increases with increasing concentration, but its value depends on the interaction. The two active motors depicted in Fig. \ref{fig:steps_results}(a) are clearly faster than a passive motor with the same monomer concentration. A comparison of the two motors with $n=3$ shows that the mean velocity does depend also on the step height $U_0$.
The fact that $v$ depends on $U_0$ is precisely the effect found for helicases by Betterton and J\"ulicher \cite{Julicher2003,Julicher2005}.

The dependence of the copying fidelity on the monomer concentration is shown in Fig. \ref{fig:steps_results}(b). The steady-growth ansatz predicts an interaction independent result, given by Eq. (\ref{eq:err_rate3}). We performed simulations for several models with different interactions between the polymerase and the junction, including a passive motor
and two active models with different three-step potentials. All show the same fidelity as a function of concentration. This fidelity is still affected by the presence of the obstacle through the kinetic coefficients $q^\pm$, and is therefore different from that of a freely propagating polymerase, which is also depicted in the figure.

\section{Discussion} \label{discussion}
Kinetic discrimination is one of the mechanisms employed by DNA polymerases and similar biological motors to increase the fidelity of copying of genetic information. This mechanism exhibits a trade-off between dissipation and fidelity. In the context of copolymerization this trade-off was studied for instance by Andrieux and Gaspard \cite{Gaspard2014}. As the system is driven further away from equilibrium, its mean velocity increases, and it attains a lower rate of copying errors.

In helicases, the mean propagation velocity can be used to deduce thermodynamic features of the motor's interaction with an obstacle \cite{Julicher2003,Julicher2005}. Experimentally, it is easier to count the copying errors in the copied strand than to follow the rate of copying.  The existence of polymerases which simultaneously copy and open single- to double-stranded junctions offers an interesting alternative. Is it possible to extract information about polymerase and junction dynamics from the enzyme's fidelity? More specifically, can one deduce whether the polymerase-junction interaction is active or passive?

The results presented in this paper help to clarify such questions. The simple model of a polymerase studied here slows down when it encounters a junction. This slowdown is indeed accompanied by an increased rate of copying errors, as can be seen from a comparison of the results of a freely propagating and non-freely propagating systems depicted in Figs. \ref{fig:HW_results}(b) and \ref{fig:steps_results}(b). This is exactly what one would anticipate in a model employing kinetic discrimination. However, our results show that the suggested correlation between the mean copying velocity and fidelity, where faster copying means better fidelity, is not always present.
Upon encountering an obstacle, the model predicts an unexpected partial decoupling between mean velocity and error rate.
The copying fidelity, as expressed by Eq. (\ref{eq:err_rate3}), is independent of details of the polymerase-junction interaction. The presence of a junction still affects the probability of copying errors, but only through the kinetic part of the transition rates (given by $q ^\pm$). In contrast, the mean velocity clearly depends on all model parameters, including the interaction. In fact, we find that active unwinding can result in higher rate of copying than that of a passive polymerase, in agreement with the results of Betterton and J\"ulicher \cite{Julicher2003,Julicher2005}.
The three non-freely propagating models in Fig. \ref{fig:steps_results} have different velocities, due to the difference between active and passive interactions. At the same time they all exhibit the same fidelity.
One still expects that when more parameters are varied, such as the monomer concentration or $q^\pm$, larger velocity will typically be accompanied by better fidelity, but it is important to point out that this is not always the case.

One may wonder whether the independence of the fidelity of the interaction is a particular property of the model studied here. Maybe the result will break down for an interaction
that is not described by a staircase potential? As explained below, the interaction dependence of fidelity found for the model is a result of the topology of the internal
state space of the system. Specifically, it emerges from the fact that this state space is one-dimensional, and therefore does not include non-trivial closed loops of
transitions. Models with different interactions would exhibit the same fidelity as long as they have a one-dimensional internal state space.

In the steady-growth ansatz, the distribution to find the polymerase at different internal states, $\Phi(j)$, becomes time independent and
furthermore satisfies Eq. (\ref{eq:phi}). This equation can be recast as
\begin{equation}
  I_{j,j-1}-I_{j+1,j}=0,
\end{equation}
where
\begin{equation}\label{eq:defI}
  I_{j+1,j} \equiv \left( \left<r^-\right>+q^+\right) \Theta^+ (j) \Phi (j) - \left( \sum_{m} r_m^+ + q^-\right) \Theta^- (j+1) \Phi (j+1)
\end{equation}
is the net flux of transitions from $j$ to $j+1$. The steady state solution for $\Phi$ must therefore satisfy
   $I_{j+1,j} =C$,
where $C$ is a $j$-independent constant.

For any reasonable model of a polymerase that unwinds a junction one expects that $\Phi (j) \rightarrow 0$ for $j\rightarrow \infty$,
expressing the fact that the polymerase and junction tend to move closer to each other. One also expects that $\Phi (j) \rightarrow 0$
for $j \rightarrow -\infty$ since otherwise the model would allow the polymerase to simply pass through the junction without unwinding it. These considerations mean that the constant $C$ must vanish, demonstrating that the detailed balance condition (\ref{eq:balance}) holds for quite general $U(j)$.
One should not take the appearance of the detailed balance
condition as evidence that the system is in thermal equilibrium. The model exhibits steady growth with a nonvanishing rate of copying, and is therefore clearly out of
equilibrium. Nevertheless, the internal state space does relax to some kind of local equilibrium.

The detailed balance condition, Eq. (\ref{eq:balance}), is the underlying reason for the interaction independent fidelity found in Sec. \ref{active}. Indeed, summation
of Eq. (\ref{eq:balance}) over $j$ results in
\begin{equation*}
\frac{\left\langle \Theta^-\right\rangle}{\left\langle \Theta^+\right\rangle }
= \rho,
\end{equation*}
for any $U(j)$ that is consistent with distributions $\Phi (j)$ that vanish at infinity.
Examination of Eq. (\ref{eq:err_rate}) shows that the mean error rate depends on $\left\langle \Theta^-\right\rangle$ and $\left\langle \Theta^+\right\rangle$
only through their ratio, $\rho$. The fact that this ratio is independent of the elastic interaction $U$ means that the mean error rate is also independent of the form of $U(j)$.

The generation of this internal detailed balance, and the resulting independence of fidelity from the elastic interaction, are an interesting manifestation of the dynamics of
copying machines. But how relevant is this phenomenon for a the few biological polymerases that remove obstacles on their own, such as reverse transcriptase? 
Can one deduce that their fidelity does not depend on the elastic interaction with a junction? Such a conclusion would be too hasty.
One should be aware that the model we constructed oversimplifies several
important aspects of the dynamics. One drastic assumption we made was to view the polymerase as a point particle. The polymerases in our cells are proteins
of a finite size. They can be squeezed by the application of an external force.

A flexible finite-sized polymerase can be modeled as an elastic spring. One can generalize the model studied here to include this aspect by considering a system
with two internal degrees of freedom, namely the size of the polymer and the distance from the edge of the polymerase to the junction. One also should include two types of
elastic interactions, a quadratic potential for the size of the polymerase, and a more general interaction between the polymerase and junction.
The crucial point is that this expanded model has an internal state space which is no longer one-dimensional. 
As a result the probability distribution in this space decays to a nonequilibrium steady state
in the steady growth regime. This is expected to lead to some degree of dependence of the fidelity on the elastic potential $U$. Further research is required
to find out whether this effect can be large.

Comparison between the model studied here and biological polymerases is further complicated by several additional factors.
We assumed a model with purely kinetic discrimination, but for instance the fidelity of reverse transcriptase is a result of a mixture of kinetic and energetic discrimination. 
We have assumed equal, constant and uniform concentrations - a condition that is unlikely to hold \textit{in vivo}.
In addition, one expects the incorporation of monomers to depend on the identity of their neighbors, leading
to correlations in the copied strand composition. All these elements must be included in the model before any biologically relevant predictions
can be made.

The approach taken here, namely studying a simplified version of the dynamics, should be viewed as a way of obtaining
qualitative understanding of copying machines, focusing on the interplay between the velocity, fidelity, and interactions with an obstacle.  It has the advantage of resulting in simple analytical expressions for observables, 
whose study can lead to qualitative insights. It is certainly worthwhile to include the additional aspects needed for a quantitative
comparison with biological copying machines. But in our view there is much to gain by first studying models in which the roles of different mechanisms can be investigated separately. 

Before concluding, we would like to point out an interesting qualitative property of the dynamics. 
Our results show that the mean velocity and the fidelity contain non-overlapping information regarding the polymerase-junction interaction and kinetics. This is clearly indicated by
the results depicted in Fig. \ref{fig:steps_results}. One should therefore strive to obtain data on both observables, and not be satisfied with measurements of only
one of them. We expect this conclusion to be rather robust, and therefore hold also for biological polymerases.

\section*{Acknowledgements}
We thank Omri Malik and Ariel Kaplan for illuminating discussions that have
initiated our interest in this topic.\\
This work was supported by the U.S.-Israel Binational Science
Foundation (Grant No. 2014405), by the Israel Science Foundation (Grant
No. 1526/15), and by the Henri Gutwirth Fund for the Promotion of
Research at the Technion.

\bibliography{polymerase}{}
\bibliographystyle{apsrev}
\end{document}